\documentclass[conference]{IEEEtran}
\IEEEoverridecommandlockouts

\usepackage{cite}
\usepackage{amsmath,amssymb,amsfonts}
\usepackage{algorithm}
\usepackage{algorithmic}
\usepackage{graphicx}
\usepackage{textcomp}
\usepackage{xcolor}
\usepackage[caption=false,font=footnotesize,labelfont=rm,textfont=rm]{subfig} 
\usepackage{stfloats}
\usepackage{url}
\usepackage{verbatim}
\usepackage{balance}
\usepackage{multirow}
\usepackage{multicol}
\usepackage{makecell}
\usepackage{tabularx}
\usepackage{booktabs}
\usepackage{titlesec}
\usepackage{amsthm}
\usepackage{bm}
\usepackage{bbm}
\usepackage{svg}

\def\BibTeX{{\rm B\kern-.05em{\sc i\kern-.025em b}\kern-.08em
    T\kern-.1667em\lower.7ex\hbox{E}\kern-.125emX}}

\newtheoremstyle{remarkbf}
  {3pt}{3pt}                 
  {\normalfont}              
  {}                         
  {\bfseries}                
  {.}{ }{}                   

\theoremstyle{remarkbf}

\begin{document}

\title{Signal Design for OTFS Dual-Functional Radar and Communications with Imperfect CSI\\
}

\author{%
Borui Du\textsuperscript{\textasteriskcentered},\;
Yumeng Zhang\textsuperscript{\textdagger},\;
Christos Masouros\textsuperscript{\textasteriskcentered},\;
Bruno Clerckx\textsuperscript{\textdaggerdbl}\\[2pt]
\textsuperscript{\textasteriskcentered} Dept. of Electronic and Electrical Engineering, University College London\\
\textsuperscript{\textdagger} Dept. of Electronic and Computer Engineering, Hong Kong University of Science and Technology\\
\textsuperscript{\textdaggerdbl} Dept. of Electrical and Electronic Engineering, Imperial College London\\
\textsuperscript{\textasteriskcentered}\{borui.du.24, c.masouros\}@ucl.ac.uk,
\textsuperscript{\textdagger}eeyzhang@ust.hk,
\textsuperscript{\textdaggerdbl}b.clerckx@imperial.ac.uk
}

\maketitle

\begin{abstract}
Orthogonal time frequency space (OTFS) offers significant advantages in managing mobility for both wireless sensing and communication systems, making it a promising candidate for dual-functional radar-communication (DFRC). However, the optimal signal design that fully exploits OTFS's potential in DFRC has not been sufficiently explored. This paper addresses this gap by formulating an optimization problem for signal design in DFRC-OTFS, incorporating both pilot-symbol design for channel estimation and data-power allocation. Specifically, we employ the integrated sidelobe level (ISL) of the ambiguity function as a radar metric, accounting for the randomness of the data symbols alongside the deterministic pilot symbols. For communication, we derive a channel capacity lower bound metric that considers channel estimation errors in OTFS. We maximize the weighted sum of sensing and communication metrics and solve the optimization problem via an alternating optimization framework. Simulations indicate that the proposed signal significantly improves the sensing-communication performance region compared with conventional signal schemes, achieving at least a 9.44 dB gain in ISL suppression for sensing, and a 4.82 dB gain in the signal-to-interference-plus-noise ratio (SINR) for communication.
\end{abstract}


\section{Introduction}
\looseness=-1
\IEEEPARstart{D}{ual-functional} radar-communication (DFRC) is a key enabler for integrated sensing and communication (ISAC) wireless systems, where a common signal is utilized to perform both sensing and communication, thereby improving hardware utilization, conserving time-frequency resources, and reducing scheduling overhead~\cite{liu_joint_2020,meng_network_level_2024}. For DFRC, waveform analysis and design are pivotal for leveraging its full potential: frequency modulated continuous wave (FMCW) offers limited communication capability~\cite{xu_radar_2023} and orthogonal frequency division multiplexing (OFDM) remains vulnerable to mobility-induced inter-carrier interference~\cite{zhang_input_2024}.

In addition to FMCW and OFDM signals, the orthogonal time frequency space (OTFS) waveform has demonstrated significant potential as a promising candidate for DFRC systems~\cite{hadani_orthogonal_2017}. Specifically, OTFS multiplexes data in the delay-Doppler (DD) domain, in which the channel is inherently sparse. By exploiting this sparsity, OTFS could effectively reconstruct the propagation environment~\cite{rasheed_sparse_2020}, particularly under large Doppler shifts. Considerable effort has been devoted to enhancing OTFS communication, particularly through pilot design. Wang et al.~\cite{wang_pilot_2021} propose a pilot design scheme for OTFS modulation by optimizing the measurement matrix structure formulated by the pilot symbols. Utilizing a heuristic optimization method, the proposed scheme significantly improves channel estimation with reduced pilot overhead. Srivastava et al.~\cite{srivastava_bayesian_2021} propose an OTFS pilot design relying on Bayesian learning-based channel estimation, which demonstrates better channel estimation performance. 

In addition to communication, early efforts have been made on OTFS signal design for radar applications~\cite{correas_serrano_emerging_2025, karimian_sichani_otfs_2025}, with some extensions to DFRC systems~\cite{song_otfs_dfrc_2023}. Specifically, Karimian-Sichani et al. optimize the ambiguity function (AF) of OTFS signals by suppressing the sidelobe level in range estimation~\cite{karimian_sichani_otfs_2025}. Furthermore, Song et al.~\cite{song_otfs_dfrc_2023} extend the signal design to DFRC scenarios by introducing phase perturbation to suppress the peak sidelobe level of the transmitted OTFS signal, but at the cost of an increased communication symbol error rate.
Additionally, Zhang et al. propose a majorization-minimization-based framework that jointly optimizes the transmitted symbols and the receiver filter in DFRC-OTFS to minimize communication interference and suppress the integrated sidelobe level (ISL) of the transmitted signal~\cite{zhang_dual_functional_2024}. This approach enhances detection performance and improves the achievable rate of the designed signal.
Very recently, Song et al.~\cite{song_low_2025} investigate OTFS signal design and optimization for ISAC systems, separating the pilot and data symbol designs for pure radar and communication functionalities and providing insightful analysis of the respective functionalities.

While previous studies have advanced our understanding of OTFS waveforms for DFRC applications, several critical issues remain unexplored, thereby preventing a thorough signal design and analysis. First, the impact of imperfect channel state information (CSI) on signal design in DFRC-OTFS has not been systematically analyzed. Second, existing research on OTFS sensing performance, particularly in terms of ISL, often fails to differentiate between the randomness of data symbols and the determinism of pilot symbols.

In this paper, we address the open issue of optimal signal design for DFRC-OTFS by jointly optimizing the communication data power and pilot symbols, thereby addressing the aforementioned problems. Specifically, we first derive a tractable channel capacity lower bound that explicitly accounts for imperfect CSI, and then formulate an ISL metric that captures the interaction between deterministic pilots and random data symbols. To the best of our knowledge, this is the first comprehensive framework for DFRC optimization in OTFS accounting for the randomness of communication symbols and channel estimation error.

\emph{Organization:}
Section~\ref{sec:sys_mod} introduces the signal and system model for OTFS. Section \ref{sec:Per_metr} derives the communication and sensing metrics for optimization in DFRC-OTFS. Section~\ref{sec:Opt_prob} outlines the optimization framework, followed by simulation verification in Section~\ref{sec:Num_res}. The paper concludes with Section~\ref{sec:conclu}.

\emph{Notations:}
$\mathbb{C}$ and $\mathbb{R}$ denote the complex and real number sets, respectively. 
$(\cdot)^{*}$ and $(\cdot)^{H}$ denote the conjugate and Hermitian operations, respectively. 
$\mathcal{CN}(0,\mathbf{C})$ denotes a zero-mean circularly symmetric complex Gaussian (CSCG) distribution with covariance $\mathbf{C}$. 
$\operatorname{vec}(\cdot)$ and $\otimes$ denote the vectorization and Kronecker product; $\operatorname{vec}^{-1}(\cdot)$ reshapes a vector into a matrix, and $\operatorname{diag}(\cdot)$ forms a diagonal matrix from a vector. 
$\varepsilon\{x\}$ denotes the expectation of the random variable $x$. $\operatorname{Re}\{x\}$ denotes the real part of the complex variable $x$.
$\mathbf{F}_{N}$ is the $N$-point normalized DFT matrix, $\mathbf{I}_{N}$ is the $N\times N$ identity matrix, $\mathbf{0}_{N}$ is the $N$-element zero vector, and $\lVert\cdot\rVert_{2}$ and $\lVert\cdot\rVert_{F}$ denote the $\ell_{2}$ and Frobenius norms.

\section{System and Signal Model}
\label{sec:sys_mod}

\begin{figure*}[ht]
    \centering
    \includegraphics[width=0.78\textwidth]{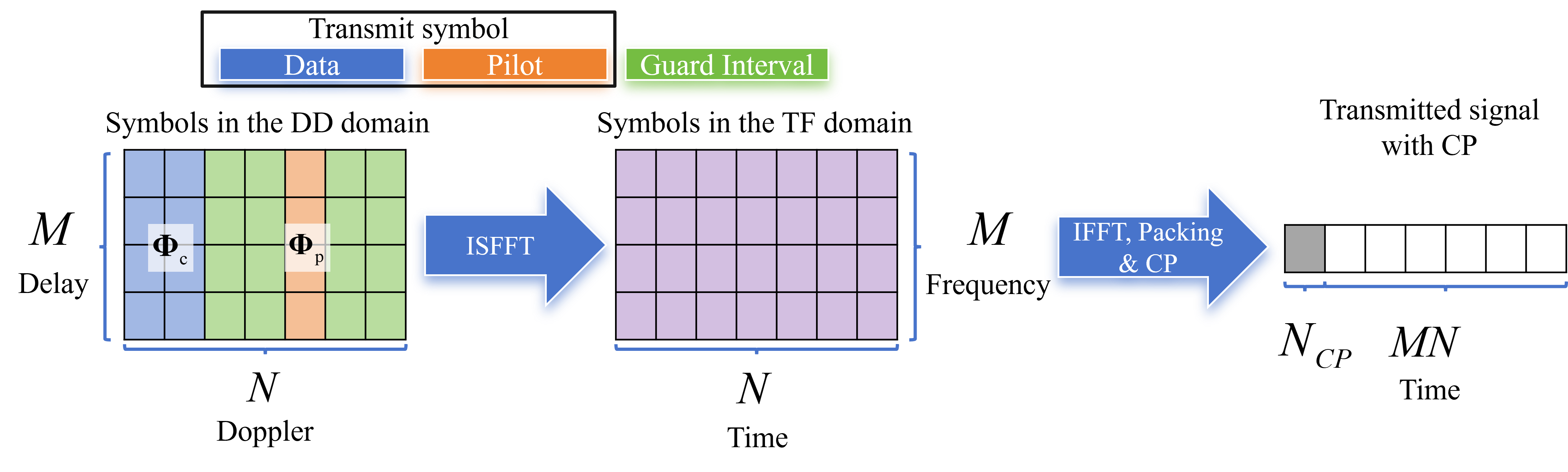}
    \caption{A toy example that illustrates the symbol arrangement (i.e., \(\mathbf{\Phi}_{\text{p}}\), \(\mathbf{\Phi}_{\text{c}}\)) and modulation process in the OTFS framework.}
    \label{fig:OTFS_mod}
\end{figure*}

\subsection{Transmitted Signal Model}
We represent the OTFS signal in the DD domain as shown in Fig.~\ref{fig:OTFS_mod}. Each frame has \(M\) subcarriers and \(N\) time slots with slot duration \(T\), and the subcarrier spacing \(\Delta f = 1/T\), yielding OTFS bandwidth \(B \triangleq M\Delta f\) and OTFS frame duration \(NT\). \(\mathbf{x}_{\text{DD}}\in\mathbb{C}^{MN\times 1}\) denotes the transmit DD-domain symbol vector. Apart from the null-symbol guard space, \(\mathbf{x}_{\text{DD}}\) comprises: (a) \(K_p\) pilot symbols \(\mathbf{x}_{\text{DD,p}}\in\mathbb{C}^{K_p\times 1}\) for channel estimation; (b) \(K_c\) data symbols \(\mathbf{x}_{\text{DD,c}}\in\mathbb{C}^{K_c\times 1}\) carrying random information. Arrangement matrices \(\mathbf{\Phi}_{\text{p}}\in\mathbb{R}^{MN\times K_p}\) and \(\mathbf{\Phi}_{\text{c}}\in\mathbb{R}^{MN\times K_c}\) map the pilot and data symbols to their designated positions within \(\mathbf{x}_{\text{DD}}\), respectively~\cite{van_der_werf_optimal_2024}.
Consequently, the OTFS transmit symbol $\mathbf{x}_\text{DD}$ is expressed as follows
\begin{align}
    \mathbf{x}_\text{DD} = \mathbf{\Phi}_\text{c}\mathbf{x}_\text{DD,c} + \mathbf{\Phi}_\text{p}\mathbf{x}_\text{DD,p}, \label{equ:arrange_matrix}
\end{align}
where the data symbols are modeled as a CSCG codebook: $\mathbf{x}_\text{DD,c} \sim \mathcal{CN}\left(0,\, p_c\mathbf{I}_{K_c}\right)$, with variance $p_c$.

\subsection{High-Mobility Communication Channel Model}
To accommodate the Doppler effect, the wireless channel is modeled as a linear time-varying system. Its effect on the OTFS signal is represented by
\begin{align}
  \mathbf{H}
    &= \sum_{i=1}^{P} \alpha_i\,\mathbf{\Pi}^{\,l_i}\,\mathbf{\Delta}^{\,k_i},
\end{align}
where $\mathbf{\Pi}=\left[\mathbf{0}^T_{MN-1}, 1; \quad\mathbf{I}_{MN-1}, \mathbf{0}_{MN-1}\right]$, and $\mathbf{\Delta}= \operatorname{diag}\!\bigl(1,\;e^{j\frac{2\pi}{MN}},\;\dots,\;e^{j\frac{2\pi(MN-1)}{MN}}\bigr)$. $P$ is the number of paths, $\alpha_i$ is the complex gain of the $i$th path, $\mathbf{\Pi}^{\,l_i}$ encodes a delay of $l_i$, and $\mathbf{\Delta}^{\,k_i}$ encodes a Doppler shift of $k_i$~\cite{raviteja_practical_2019}. $L$ and $Q$ denote the maximum delay and Doppler indices, i.e., $l_i \in \left\{0, 1, \cdots, L\right\}$ and $k_i \in \left\{0, 1, \cdots, Q\right\}$, respectively.

\subsection{Received Signal Model at the Communication Receiver}
After propagating through the wireless channel, the received DD-domain signal at the communication receiver is
\begin{subequations}
\begin{align}
  \mathbf{y}_{\text{DD}}
    &= \mathbf{H}_{\text{DD}}\,\mathbf{x}_{\text{DD}} + \mathbf{n}_{\text{DD}}, \label{equ:OTFS_sys} \\[2pt]
  \text{with}\quad
  \mathbf{H}_{\text{DD}}
    &\triangleq \bigl(\mathbf{F}_N \otimes \mathbf{I}_M\bigr)\,\mathbf{H}\,\bigl(\mathbf{F}_N^{H} \otimes \mathbf{I}_M\bigr), \label{equ:h2H}
\end{align}
\end{subequations}
where $\mathbf{n}_{\text{DD}}\in\mathbb{C}^{MN\times 1}$ and $\mathbf{n}_{\text{DD}}\sim \mathcal{CN}\!\left(\mathbf{0},\,\sigma_{\mathbf{n}}^{2}\mathbf{I}_{MN}\right)$ is the additive white Gaussian noise (AWGN), and $\mathbf{H}_{\text{DD}}$ is the effective channel that captures wireless propagation in the DD domain.

As counterparts to $\mathbf{\Phi}_{\text{p}}$ and $\mathbf{\Phi}_{\text{c}}$ in (\ref{equ:arrange_matrix}), we adopt selection matrices
$\mathbf{\Psi}_{\text{p}}\in\mathbb{R}^{MN\times R_{\text{p}}}$ and
$\mathbf{\Psi}_{\text{c}}\in\mathbb{R}^{MN\times R_{\text{c}}}$ applied to the received signal to extract pilot and data symbols from the DD domain. To decouple pilots and data, we assume a guard interval (GI) that prevents mutual interference~\cite{van_der_werf_optimal_2024}.

Therefore, the received data symbols could be written as
\begin{align}
  \mathbf{y}_{\text{DD,c}}
    &= \mathbf{\Psi}_{\text{c}}^{H}\mathbf{y}_{\text{DD}}
     = \mathbf{\Psi}_{\text{c}}^{H}\mathbf{H}_{\text{DD}}
       \bigl(\mathbf{\Phi}_{\text{c}}\mathbf{x}_{\text{DD,c}}+\mathbf{\Phi}_{\text{p}}\mathbf{x}_{\text{DD,p}}\bigr)
       + \mathbf{\Psi}_{\text{c}}^{H}\mathbf{n}_{\text{DD}} \\
    &\triangleq \mathbf{H}_{\text{c}}\mathbf{x}_{\text{DD,c}} + \mathbf{n}_{\text{c}}, \label{equ:OTFS_sys_data}
\end{align}
where $\mathbf{H}_{\text{c}}\triangleq \mathbf{\Psi}_{\text{c}}^{H}\mathbf{H}_{\text{DD}}\mathbf{\Phi}_{\text{c}}$ and
$\mathbf{n}_{\text{c}}\triangleq \mathbf{\Psi}_{\text{c}}^{H}\mathbf{n}_{\text{DD}}$. 
Herein, $\mathbf{y}_{\text{DD,c}}\in\mathbb{C}^{R_{\text{c}}\times 1}$, and $\mathbf{n}_{\text{c}}$ is the AWGN in the data-symbol space. 
The cross term $\mathbf{\Psi}_{\text{c}}^{H}\mathbf{H}_{\text{DD}}\mathbf{\Phi}_{\text{p}}\mathbf{x}_{\text{DD,p}}$\,=\,0, as guaranteed by the GI assumption.

Similarly, the received pilot symbols are given by
\begin{subequations}
\begin{align}
  \mathbf{y}_{\text{DD,p}}
    &= \mathbf{\Psi}_{\text{p}}^{H}\mathbf{H}_{\text{DD}}\mathbf{\Phi}_{\text{p}}\mathbf{x}_{\text{DD,p}}
     + \mathbf{\Psi}_{\text{p}}^{H}\mathbf{n}_{\text{DD}}, \label{equ:pilot_DD}\\
    &= \mathbf{\Omega}_{\text{DD,p}}\,\mathbf{h} + \mathbf{n}_{\text{p}}, \label{equ:pilot_Omega_h}\\
      \text{with }\;
      \mathbf{\Omega}_{\text{DD,p}}
      &= \bigl[\,\bm{\omega}^{0}_{0,\text{DD,p}},\;\bm{\omega}^{0}_{1,\text{DD,p}},\;\dots,\;\bm{\omega}^{Q}_{L,\text{DD,p}}\,\bigr]
         \in \mathbb{C}^{R_{\text{p}}\times K_h},\\
      \text{and }\;\bm{\omega}^{j}_{i,\text{DD,p}}
      &= \mathbf{\Psi}_{\text{p}}^{H}\,(\mathbf{F}_{N}\!\otimes\!\mathbf{I}_{M})\,
        \mathbf{\Pi}^{i}\mathbf{\Delta}^{j}\,
        (\mathbf{F}_{N}^{H}\!\otimes\!\mathbf{I}_{M})\,
        \mathbf{\Phi}_{\text{p}}\,\mathbf{x}_{\text{DD,p}},
\end{align}
\end{subequations}
where \(\mathbf{y}_{\text{DD,p}}\in\mathbb{C}^{R_{\text{p}}\times 1}\), and \(\mathbf{n}_{\text{p}}\triangleq \mathbf{\Psi}_{\text{p}}^{H}\mathbf{n}_{\text{DD}}\) is the corresponding noise. $\mathbf{h}\in\mathbb{C}^{K_h\times 1}$ with $K_h \triangleq (L+1)(Q+1)$ collects the DD channel coefficients. Each entry of $\mathbf{h}$ follows a CSCG distribution, i.e., it is nonzero with probability $p$, and when nonzero, follows CSCG as $\mathcal{CN}(0,\sigma_{\mathbf{h}}^{2})$; otherwise, it is $0$ with probability $1-p$.  
The pilot dictionary columns $\bm{\omega}^{j}_{i,\text{DD,p}}\in\mathbb{C}^{R_{\text{p}}\times 1}$ encode the pilot placement ($\mathbf{\Phi}_{\text{p}},\mathbf{\Psi}_{\text{p}}$), the pilot symbols ($\mathbf{x}_{\text{DD,p}}$), and the $(i,j)$ delay-Doppler shifts.

\section{Performance Metrics}
\label{sec:Per_metr}

\subsection{Channel Capacity Lower Bound}
Given (\ref{equ:pilot_Omega_h}), the linear minimum mean square error (LMMSE) channel estimate could be expressed as~\cite{srivastava_bayesian_2021}
\begin{equation}
    {\widehat{\mathbf{h}}}_\text{LMMSE}
    = \left(\mathbf{\Omega}^H_\text{DD,p}\mathbf{C}^{-1}_{\mathbf{n}_\text{p}}{\mathbf{\Omega}}_\text{DD,p} + \mathbf{C}^{-1}_\mathbf{h}\right)^{-1}
      {\mathbf{\Omega}}^H_\text{DD,p}\mathbf{C}^{-1}_{\mathbf{n}_\text{p}}\mathbf{y}_\text{DD,p},
\end{equation}
where \( \mathbf{C}_{\mathbf{n}_\text{p}} \) is the noise covariance matrix of the pilot space. The estimated channel matrix \( \widehat{\mathbf{H}}_\text{DD} \) is then recovered from (\ref{equ:h2H}) using ${\widehat{\mathbf{h}}}_\text{LMMSE}\in\mathbb{C}^{K_h \times 1}$. Also, we denote \( \widehat{\mathbf{H}}_\text{c} \triangleq \mathbf{\Psi}^H_\text{c} \widehat{\mathbf{H}}_\text{DD}\mathbf{\Phi}_\text{c}\) and \( \widehat{\mathbf{H}}_\text{p} \triangleq \mathbf{\Psi}^H_\text{p} \widehat{\mathbf{H}}_\text{DD}\mathbf{\Phi}_\text{p}\) as the estimated channel matrices for data and pilot symbols, respectively. Therefore, $\mathbf{y}_\text{DD,c}$ could be rewritten as follows
\begin{align}
     \mathbf{y}_\text{DD,c} &= \widehat{\mathbf{H}}_\text{c}\mathbf{x}_\text{DD,c} + \left(\mathbf{H}_\text{c} -\widehat{\mathbf{H}}_\text{c}\right)\mathbf{x}_\text{DD,c} + \mathbf{n}_\text{c} \\ 
     &= \widehat{\mathbf{H}}_\text{c}\mathbf{x}_\text{DD,c} + \mathbf{v}_\text{DD,c}, \label{equ:OTFS_sys_data_est_error}
\end{align}
where $\mathbf{v}_\text{DD,c}\triangleq\left(\mathbf{H}_\text{c} -\widehat{\mathbf{H}}_\text{c}\right)\mathbf{x}_\text{DD,c} + \mathbf{n}_\text{c}$ is the effective noise vector that accounts for channel estimation error and AWGN. Based on the received signal model in (\ref{equ:OTFS_sys_data_est_error}), the channel capacity lower bound for OTFS is given by~\cite{xiaoli_ma_optimal_2003}
\begin{figure*}
    \begin{align}
        \mathbf{C}_{\mathbf{v}_\text{DD,c}}
        &\mathop \preceq \Bigg[ 
        p_c \operatorname{Tr} \Bigg( 
        p \sigma_{\mathbf{h}}^2 
        \left( 
        \mathbf{I}_{K_h} + 
        \frac{p \sigma_{\mathbf{h}}^2}{\sigma_{\mathbf{n}}^2} 
        \bm{\Omega}^H_\text{DD,p} \bm{\Omega}_\text{DD,p} 
        \right)^{-1} 
        \Bigg) 
        + \sigma^2_\mathbf{n} 
        \Bigg] \mathbf{I}_{R_c}
        \label{equ:tr_app_result_specific}
    \end{align}
    \hrule
\end{figure*}

\begin{equation}
    C \ge \frac{f_{\text{CP}}}{MN}\, \varepsilon\!\left[\log \det \!\left(\mathbf{I}_{R_c} + p_c\, \mathbf{C}^{-1}_{\mathbf{v}_\text{DD,c}} \,\widehat{\mathbf{H}}_\text{c} \,\widehat{\mathbf{H}}^{H}_\text{c}\right)\right], \label{equ:OTFS_channel_capacity_lower_bound}
\end{equation}
where \( f_{\text{CP}} \triangleq \frac{MN}{MN+N_\text{CP}} \) accounts for the reduced-CP overhead~\cite{raviteja_practical_2019}; \(N_\text{CP}\) denotes the length of the reduced-CP, which is assumed to be larger than the maximum channel delay. \(\mathbf{C}_{\mathbf{v}_\text{DD,c}} = \varepsilon\!\left[\mathbf{v}_\text{DD,c}\mathbf{v}_\text{DD,c}^H\right]\) denotes the covariance of \(\mathbf{v}_\text{DD,c}\), and is detailed in~(\ref{equ:tr_app_result_specific}), whose derivation is omitted for space reasons~\cite{van_der_werf_optimal_2024}. Consequently, we obtain (\ref{equ:OTFS_channel_capacity_lower_bound_scalar}) as a tractable channel capacity lower bound.
\begin{figure*}[ht]
    \begin{equation}
        C \ge \frac{f_\text{CP}}{MN} \varepsilon \left\{\log \det \left[\mathbf{I}_{R_c} + \underbrace{\frac{p_c}{\sigma^2_{\mathbf{n}}} \left[ \frac{p_c}{\sigma^2_{\mathbf{n}}} \operatorname{Tr} \left( p\sigma_{\mathbf{h}}^2 \left(\mathbf{I}_{K_{h}} + \frac{p\sigma_{\mathbf{h}}^2}{\sigma_{\mathbf{n}}^2}{{\bm{\Omega}}^H _\text{DD,p}}{\bm{\Omega}} _\text{DD,p}\right)^{-1}\right) + 1\right]^{-1}}_\text{Scalar} \mathbf{\widehat H}_\text{c} \mathbf{\widehat H}^H_\text{c}\right]\right\} \label{equ:OTFS_channel_capacity_lower_bound_scalar}
    \end{equation}
    \hrule
\end{figure*}
\begin{figure*}[ht]
\begin{align}
    \text{SINR} &\triangleq 
    \frac{p_c}{\sigma^2_{\mathbf{n}}} \left[ \frac{p_c}{\sigma^2_{\mathbf{n}}} \operatorname{Tr} \left( p\sigma_{\mathbf{h}}^2 \left(\mathbf{I}_{K_{h}} + \frac{p\sigma_{\mathbf{h}}^2}{\sigma_{\mathbf{n}}^2}{{\bm{\Omega}}^H _\text{DD,p}}{\bm{\Omega}} _\text{DD,p}\right)^{-1}\right) + 1\right]^{-1}
\label{equ:opt_SINR}
\end{align}
\hrule
\end{figure*} 

From \eqref{equ:OTFS_channel_capacity_lower_bound_scalar}, the scalar (as marked) inside the logarithm dominates the capacity lower bound once channel estimation error is considered. We define this scalar term as the signal-to-interference-plus-noise-ratio (SINR) in \eqref{equ:opt_SINR} \cite{van_der_werf_optimal_2024}, and use it as the communication metric. Moreover, \(\text{SINR}\) is concave in \(p_c\) and is optimization friendly.

\subsection{Integrated Sidelobe Level of the Ambiguity Function}
In terms of the radar metric, we focus on the ISL properties of the AF of the transmitted signal, which dominate the sensing performance. The cross-correlation of a target at delay bin $l$ and Doppler bin $k$ could be written as $f_{lk} = \mathbf{x}_\text{DD}^H \,\mathbf{A}_{lk}\,\mathbf{x}_\text{DD}$,
where 
\begin{align}
    \mathbf{A}_{lk}
    &= \bigl(\mathbf{F}_{N}\otimes \mathbf{I}_{M}\bigr)\,
       \mathbf{\Gamma}^H\,\mathbf{J}_{l}\,\mathbf{D}_{k}\,\mathbf{\Gamma}\,
       \bigl(\mathbf{F}_{N}^H\otimes \mathbf{I}_{M}\bigr)
\end{align} is a constant matrix. Moreover, \( \mathbf{\Gamma} \in \mathbb{R}^{(MN + N_{\text{CP}}) \times MN} \) is the CP arrangement matrix, which is given by~\cite{zhang_dual_functional_2024}. \(\mathbf{J}_{l}\in\mathbb{C}^{(MN+N_{\text{CP}})\times(MN+N_{\text{CP}})}\) is the linear delay matrix with entries \([\mathbf{J}_{l}]_{m,n}=1\) if \(n=m-l\) and \(l< m\le MN+N_{\text{CP}}\), and \(0\) otherwise. $\mathbf{J}_{-l} = \mathbf{J}^T_{l}$. \(\mathbf{D}_{k} = \operatorname{diag}\bigl(1,\, e^{-\frac{j2\pi k}{MN + N_\text{CP}}},\, \ldots,\, e^{-\frac{j2\pi (MN + N_\text{CP}-1)k}{MN + N_\text{CP}}}\bigr)\) is the Doppler-shift matrix for the received signal. Thus, after simple algebra, the expected ISL is expressed as
\begin{subequations}
    \label{equ:ISL}
    \begin{align}
    \text{ISL} &= \sum\limits_{l,k} \Bigg[\mathbf{x}_\text{DD,p}^H \mathbf{A}_{\text{p},lk} \mathbf{x}_\text{DD,p} \mathbf{x}_\text{DD,p}^H \mathbf{A}_{\text{p},lk}^H \mathbf{x}_\text{DD,p}
     \notag\\
    &\quad\quad\quad\;+p_c^2 \big( a_{lk} + \left| b_{lk} \right|^2 \big) + p_c \mathbf{x}_\text{DD,p}^H \mathbf{B}_{lk} \mathbf{x}_\text{DD,p} \notag\\
    &\quad\quad\quad\;+ 2p_c \operatorname{Re} \big( b_{lk} \mathbf{x}_\text{DD,p}^H \mathbf{A}_{\text{p},lk}^H \mathbf{x}_\text{DD,p} \big) \Bigg], \label{equ:ISL_simp}\\
    \text{with }&\mathbf{B}_{lk} \triangleq \mathbf{A}_{\text{p,c},lk} \mathbf{A}_{\text{c,p},lk} + \mathbf{A}_{\text{p,c},lk} \mathbf{A}_{\text{p,c},lk}^H  + \mathbf{A}_{\text{c,p},lk}^H \mathbf{A}_{\text{c,p},lk} \notag \\ 
    &\quad\quad\;\;+ \mathbf{A}_{\text{c,p},lk}^H \mathbf{A}_{\text{p,c},lk}^H,
    \end{align}
\end{subequations} where \(l\in\{-\widehat{L},\ldots,\widehat{L}\}\) and \(k\in\{-\widehat{Q},\ldots,\widehat{Q}\}\) denote the candidate delay and Doppler bins used in the ISL calculation, and \(a_{lk}\triangleq \operatorname{Tr}\!\big(\mathbf{\Phi}_{\text{c}}^{H}\mathbf{A}_{lk}\mathbf{\Phi}_{\text{c}}\mathbf{\Phi}_{\text{c}}^{H}\mathbf{A}_{lk}^{H}\mathbf{\Phi}_{\text{c}}\big)\),
\(b_{lk}\triangleq \operatorname{Tr}\!\big(\mathbf{\Phi}_{\text{c}}^{H}\mathbf{A}_{lk}\mathbf{\Phi}_{\text{c}}\big)\),
\(\mathbf{A}_{\text{p},lk}\triangleq \mathbf{\Phi}_{\text{p}}^{H}\mathbf{A}_{lk}\mathbf{\Phi}_{\text{p}}\),
\(\mathbf{A}_{\text{p,c},lk}\triangleq \mathbf{\Phi}_{\text{p}}^{H}\mathbf{A}_{lk}^{H}\mathbf{\Phi}_{\text{c}}\), and
\(\mathbf{A}_{\text{c,p},lk}\triangleq \mathbf{\Phi}_{\text{c}}^{H}\mathbf{A}_{lk}^{H}\mathbf{\Phi}_{\text{p}}\).

\section{Optimization for Signal Design}
\label{sec:Opt_prob}
In this section, we formulate a joint DFRC-OTFS optimization problem, with respect to (w.r.t.) the pilot symbols ($\mathbf {x} _\text{DD,p}$) and the data-symbol power ($p_c$), to maximize the weighted sum of negative ISL and SINR, thereby illuminating the sensing-communication trade-off in the following analysis and simulations. The problem is solved via an alternating optimization (AO) framework, whose second subproblem is solved using an inner alternating direction method of multipliers (ADMM)-successive convex approximation (SCA).

\subsection{Problem Formulation}
\label{sec:Opt_prob:Prob_formulation}
Given the defined $\text{SINR}$ in (\ref{equ:opt_SINR}) and $\text{ISL}$ in (\ref{equ:ISL_simp}), we first formulate the DFRC-OTFS optimization problem as follows
\begin{subequations} \label{equ:P1_OBJ}
\begin{align}
    \mathop {\max }\limits_{p_c,{\mathbf {x} _\text{DD,p}}} \quad 
    & \eta \,\text{SINR}\left(p_c,{\mathbf {x} _\text{DD,p}}\right) - \overline{\eta}\text{ISL}\left(p_c,{\mathbf {x} _\text{DD,p}}\right), \label{equ:objective_sub} \\
    \text{s.t.} \quad 
    & P_{\text{T}}\left(p_c,{\mathbf {x} _\text{DD,p}}\right) \leq {P_\text{max}}, \label{equ:P1_C1} \\
    & \overline{f}_{00}\left(p_c,{\mathbf {x} _\text{DD,p}}\right) \geq \xi_{\text{min}} , \label{equ:P1_C2}
\end{align}
\end{subequations}
\begin{subequations}
\begin{align}
    \text{with }P_{\text{T}}\left(p_c,{\mathbf {x} _\text{DD,p}}\right) &\buildrel\Delta\over= \frac{1}{MN+N_{\text{CP}}} \overline{f}_{00}\left(p_c,{\mathbf {x} _\text{DD,p}}\right),\\
    \text{with }\overline{f}_{00}\left(p_c,{\mathbf {x} _\text{DD,p}}\right) &\buildrel\Delta\over= \varepsilon\left\{{f}_{00}\right\}\\
    &=p_c\operatorname{Tr}\big( 
        {{\bm{\Phi}}_{\text{c}}^H {\bf{B}}^H {\bf{B}} {{\bm{\Phi}}_{\text{c}}}} 
        \big) \nonumber \\ &\quad +
        {{\mathbf{x}}_{{\text{DD,p}}}^H{\bm{\Phi}}_{\text{p}}^H {\bf{B}}^H {\bf{B}} 
        {\bm{\Phi}}_{\text{p}}{\mathbf{x}}_{{\text{DD,p}}}},
\end{align}
\end{subequations}
\(\eta\) is the ISAC trade-off weight and \(\overline{\eta}\triangleq 1-\eta\).
Constraint \eqref{equ:P1_C1} enforces the transmit power budget, where \(P_{\text{T}}\) is the actual emitted power.
Constraint \eqref{equ:P1_C2} imposes a mainlobe requirement on the AF to ensure a clear mainlobe-sidelobe gap for radar, with
\(\mathbf{B}\triangleq \mathbf{\Gamma}\bigl(\mathbf{F}_{N}^{H}\otimes \mathbf{I}_{M}\bigr)\).
In \eqref{equ:P1_C2}, \(\overline{f}_{00}\!\left(p_c,\mathbf{x}_{\text{DD,p}}\right)\) denotes the mainlobe of the expected AF of the transmit signal, and \(\xi_{\min}\) is the minimum acceptable mainlobe level.

\subsection{Alternating Optimization}
\label{sec:Opt_prob:Alter_opt}
We solve \eqref{equ:P1_OBJ} by reformulating it into convex subproblems. 
The communication term \(\text{SINR}\) is non-concave and involves costly matrix inversions. 
To address this, we introduce an auxiliary variable \(s_{1}\) for \eqref{equ:opt_SINR} to simplify the inversion.
We define
\begin{equation}
\begin{split}
  \text{SINR}'
  \bigl(p_c,s_{1}\bigr)
  &= \frac{p_c}{\sigma^{2}_{\mathbf{n}}}
     \Bigl(\frac{p_c}{\sigma^{2}_{\mathbf{n}}} s_{1}+1\Bigr)^{-1}.
\end{split}
\label{equ:SINR_s1}
\end{equation}
The equivalence is maintained by including the following inequality constraint
\begin{equation}
\begin{split}
  s_{1}\;\ge\; \operatorname{Tr}\!\Bigl(
    p\,\sigma^{2}_{\mathbf{h}}\bigl(\mathbf{I}_{K_h}
    &+ \frac{p\,\sigma^{2}_{\mathbf{h}}}{\sigma^{2}_{\mathbf{n}}}\,
      {\mathbf{\Omega}}_{\text{DD,p}}^{H}{\mathbf{\Omega}}_{\text{DD,p}}
    \bigr)^{-1}
  \Bigr).
\end{split}
\label{equ:s1_const}
\end{equation}
The problem in \eqref{equ:P1_OBJ} is thus reformulated as
\begin{subequations} \label{equ:P2}
\begin{align}
    \mathop {\max }\limits_{p_c, s_1, {\mathbf{x}_\text{DD,p}}} \quad 
    & \eta \,{\text{SINR}^\prime}\left(p_c,s_1\right) - \overline{\eta}\text{ISL}\left(p_c,{\mathbf {x} _\text{DD,p}}\right), \label{equ:P2_objective} \\
    \text{s.t.} \quad 
    & (\ref{equ:P1_C1}), (\ref{equ:P1_C2}), \text{ and } (\ref{equ:s1_const}). \label{equ:P2_C}
\end{align}
\end{subequations}

From \eqref{equ:P2}, \(p_c\) appears only in constraints \eqref{equ:P1_C1}-\eqref{equ:P1_C2} within \eqref{equ:P2_C}, and the objective is concave in \(p_c\) for fixed \((s_{1},\mathbf{x}_{\text{DD,p}})\). 
We therefore adopt an AO scheme to exploit this concavity. 
The resulting subproblems are given below, where \(n\) denotes the AO iteration index.

\emph{1) AO Subproblem 1:} This subproblem optimizes the power of the data symbols, \(p_c\), given fixed $s_{1}^{(n)}$ and ${\mathbf{x}}^{(n)}_{{\text{DD}},{\text{p}}}$, which could be formulated as follows
\begin{subequations} \label{equ:SP1_simp}
\begin{align}
\max_{p_c}\;&\eta\,\text{SINR}'\bigl(p_c,s_1^{(n)}\bigr)-\overline{\eta}\,\text{ISL}\bigl(p_c,\mathbf{x}_\text{DD,p}^{(n)}\bigr)\label{equ:SP1_simp_objective}\\
\text{s.t.}\;&P_\text{T}\bigl(p_c,\mathbf{x}_\text{DD,p}^{(n)}\bigr)\le P_\text{max},\label{equ:SP1_simp_C1}\\
&\overline{f}_{00}\bigl(p_c,\mathbf{x}_\text{DD,p}^{(n)}\bigr) \ge \xi_\text{min},\label{equ:SP1_simp_C2}
\end{align}
\end{subequations}
which is a convex problem and could be solved using convex optimization.

\emph{2) AO Subproblem 2:} 
The second AO subproblem is non-convex, particularly w.r.t. ${\mathbf{x}}_{\text{DD},\text{p}}$. To this end, we replace \( {\mathbf{x}}_{\text{DD},\text{p}} \) with \( {\mathbf{x}}_{\text{DD},\text{p,1}} \) and \( {\mathbf{x}}_{\text{DD},\text{p,2}}\) in all equations to promote convexity, and we introduce a constraint \( {\mathbf{x}}_{\text{DD},\text{p,1}} = {\mathbf{x}}_{\text{DD},\text{p,2}} \) to ensure consistency. In addition, a slack matrix variable $\mathbf{A}$ is introduced to further relax constraint (\ref{equ:P2_C}) and handle the matrix inverse. Specifically, the original optimization subproblem is reformulated as follows
\begin{subequations} \label{equ:SP3}
\begin{align}
    \mathop{\min}\limits_{\substack{s_{1},{{\mathbf{x}}_{\text{DD},\text{p,1}}},\\{\mathbf{A}},{{\mathbf{x}}_{\text{DD},\text{p,2}}}}} \quad
    &\! -\eta\,\text{SINR}'\left(p_{c}^{(n+1)},\,s_1\right) \nonumber \\
    &\! +\overline\eta\,\text{ISL}'\left(p_{c}^{(n+1)},\,\mathbf{x}_{\text{DD,p,1}},\,\mathbf{x}_{\text{DD,p,2}}\right) \nonumber \\
    &\! + \frac{\rho}{2} \Big\| 
        {{\mathbf{x}}_{{\text{DD}},{\text{p,1}}}} 
        - {{\mathbf{x}}_{{\text{DD}},{\text{p,2}}}} 
        + \mathbf{d} 
    \Big\|^2_2 \nonumber \\
    &\! + \frac{\zeta}{2} \Big\|
        \mathbf{A}\,\mathbf{\Xi}\!\left(\mathbf{x}_{\text{DD},\text{p,1}},\mathbf{x}_{\text{DD},\text{p,2}}\right) - \mathbf{I}
    \Big\|^2_F, \label{equ:SP3_objective}\\
    \text{s.t.} \quad
    & s_{1} \ge \operatorname{Tr}\left(p\sigma_{\mathbf{h}}^2 \mathbf{A}\right), \label{equ:SP3_C1}\\
    & P_{\text{T}}\left(p_c^{(n+1)},{\mathbf {x} _\text{DD,p,1}},{\mathbf {x} _\text{DD,p,2}}\right) \le P_\text{max}, \label{equ:SP3_C2}\\
    & \overline{f}_{00}\left(p_c^{(n+1)},{\mathbf {x} _\text{DD,p,1}},{\mathbf {x} _\text{DD,p,2}}\right) \ge \xi_\text{min}, \label{equ:SP3_C3}\\
    & \mathbf{x}_{\text{DD},\text{p,1}} = \mathbf{x}_{\text{DD},\text{p,2}}, \label{equ:SP3_C4}
\end{align}
\end{subequations}
\begin{align}
\text{with }&\;\text{ISL}'\!\left(p_c,
                    \mathbf{x}_{\text{DD,p,1}},
                    \mathbf{x}_{\text{DD,p,2}}\right)\notag\\
  \triangleq &\sum_{l,k} \Bigl[
      p_c^2\!\left(a_{lk} + \lvert b_{lk}\rvert^{2}\right) 
  + p_c
        \mathbf{x}_{\text{DD,p,2}}^{\!H}
        \mathbf{B}_{lk}
        \mathbf{x}_{\text{DD,p,1}} \notag\\
  &+ 2p_c
        \operatorname{Re}\!\left(
          b_{lk}\,
          \mathbf{x}_{\text{DD,p,2}}^{\!H}
          \mathbf{A}_{\text{p},lk}^{H}
          \mathbf{x}_{\text{DD,p,1}}
        \right) \notag\\
  &+ \mathbf{x}_{\text{DD,p,1}}^{\!H}
        \mathbf{A}_{\text{p},lk}
        \mathbf{x}_{\text{DD,p,2}}\,
        \mathbf{x}_{\text{DD,p,2}}^{\!H}
        \mathbf{A}_{\text{p},lk}^{H}
        \mathbf{x}_{\text{DD,p,1}}
    \Bigr],
\label{equ:ISL_simp_ADMM}
\end{align}
where \(P_{\text{T}}\!\left(p_c,\mathbf{x}_{\text{DD,p,1}},\mathbf{x}_{\text{DD,p,2}}\right)\) is obtained by replacing the first and second occurrences of \(\mathbf{x}_{\text{DD,p}}\) in \(P_{\text{T}}\) with \(\mathbf{x}_{\text{DD,p,1}}\) and \(\mathbf{x}_{\text{DD,p,2}}\), respectively; \(\overline{f}_{00}\!\left(p_c,\mathbf{x}_{\text{DD,p,1}},\mathbf{x}_{\text{DD,p,2}}\right)\) is defined analogously. $\mathbf{\Xi}\!\left(\mathbf{x}_{\text{DD},\text{p,1}},\mathbf{x}_{\text{DD},\text{p,2}}\right) \triangleq \mathbf{I}_{K_h} + \frac{p\sigma_{\mathbf{h}}^2}{\sigma_{\mathbf{n}}^2} \left(\mathbf{I}_{K_h}\otimes\mathbf{x}_{\text{DD},\text{p,2}}^H\right)\widetilde{\mathbf{\Omega}}_{\text{DD,p}}^H \widetilde{\mathbf{\Omega}}_{\text{DD,p}} \left(\mathbf{I}_{K_h}\otimes\mathbf{x}_{\text{DD},\text{p,1}}\right)$. \(\widetilde{\mathbf{\Omega}}_{\text{DD,p}}\) denotes the extended pilot dictionary, defined as $\widetilde{\mathbf{\Omega}}_\text{DD,p} = \left[ \widetilde{\mathbf{\Omega}}_{0,\text{DD,p}}^0, \widetilde{\mathbf{\Omega}}_{1,\text{DD,p}}^{0}, \cdots, \widetilde{\mathbf{\Omega}}_{{L},\text{DD,p}}^{Q} \right]$, and ${\widetilde{\mathbf{\Omega}}}_{i,\text{DD,p}}^j = \mathbf{\Psi}^H_\text{p}\left( {\mathbf{F}_N \otimes {\mathbf{I}_M}} \right)\mathbf{\Pi} ^i\mathbf{\Delta}^j\left( {{\mathbf{F}_N^H} \otimes {\mathbf{I}_M}} \right)\mathbf{\Phi}_\text{p}$. Besides, ${\mathbf{\Omega}}_{\text{DD,p}} = \widetilde{\mathbf{\Omega}}_{\text{DD,p}}\left(\mathbf{I}_{K_h} \otimes \mathbf{x}_\text{DD,p}\right)$. $\rho$ and $\zeta$ are penalty coefficients for the ADMM framework and the slackness operation, respectively. After introducing \( \mathbf{x}_{\text{DD,p,1}} \) and \( \mathbf{x}_{\text{DD,p,2}} \), the ISL expression is updated to~(\ref{equ:ISL_simp_ADMM}), which is separately convex w.r.t. $\mathbf{x}_\text{DD,p,1}$ and $\mathbf{x}_\text{DD,p,2}$ after the ADMM reformulation.

Through ADMM, we iteratively update \(s_{1}\), \(\mathbf{x}_{\text{DD,p,1}}\), \(\mathbf{x}_{\text{DD,p,2}}\), and \(\mathbf{A}\) until they converge. Specifically, at AO iteration \(n\) and ADMM iteration \(m\), the problem~\eqref{equ:SP3} is updated as follows
\begin{align}
\begin{bmatrix}
    s^{(n,m+1)}_{1}\\
    \mathbf{x}^{(n,m+1)}_{\text{DD,p,1}}
\end{bmatrix}
:=& \arg\min_{\substack{
            s_{1}, \\ 
            {{\mathbf{x}}_{{\text{DD}},{\text{p,1}}}}
        }}
\biggl\{
    -\eta\,
    \text{SINR}'\left(p_c^{(n+1)}, s_1\right) \nonumber\\
&+\overline{\eta}\,
    \text{ISL}'\left(p_c^{(n+1)}, {{\mathbf{x}}_{{\text{DD}},{\text{p,1}}}}, {{\mathbf{x}}^{(n,m)}_{{\text{DD}},{\text{p,2}}}}\right)\, \nonumber \\ 
&+ \frac{\rho}{2} \Big\| 
        {{\mathbf{x}}_{{\text{DD}},{\text{p,1}}}} 
        - {{\mathbf{x}}^{(n,m)}_{{\text{DD}},{\text{p,2}}}} 
        + \mathbf{d}^{(n,m)} 
    \Big\|^2_2 \nonumber \\
&+ \frac{\zeta}{2} \Big\|
    \mathbf{A}^{(n,m)} \mathbf{\Xi}\left(\mathbf{x}_{\text{DD},\text{p,1}},\mathbf{x}^{(n,m)}_{\text{DD},\text{p,2}}\right) - \mathbf{I}
\Big\|^2_F\nonumber \\
& + g_1 + g_2 + g_3\biggr\}
,\label{equ:ADMM_x1}\\
\mathbf{x}^{(n,m+1)}_{\text{DD,p,2}}
:=& \nonumber \\ 
\arg\min_{\substack{
            {{\mathbf{x}}_{{\text{DD}},{\text{p,2}}}}
        }}
\biggl\{&\;\overline{\eta}\,
    \text{ISL}'\left(p_c^{(n+1)}, {{\mathbf{x}}^{(n,m+1)}_{{\text{DD}},{\text{p,1}}}}, {{\mathbf{x}}_{{\text{DD}},{\text{p,2}}}}\right)
\, \nonumber \\ 
&+ \frac{\rho}{2} \Big\| 
        {{\mathbf{x}}^{(n,m+1)}_{{\text{DD}},{\text{p,1}}}} 
        - {{\mathbf{x}}_{{\text{DD}},{\text{p,2}}}} 
        + \mathbf{d}^{(n,m)} 
    \Big\|^2_2 \nonumber \\
&+ \frac{\zeta}{2} \Big\|
    \mathbf{A}^{(n,m)} \mathbf{\Xi}\left(\mathbf{x}^{(n,m+1)}_{\text{DD},\text{p,1}},\mathbf{x}_{\text{DD},\text{p,2}}\right) - \mathbf{I}
\Big\|^2_F\nonumber \\
& + h_1 + h_2 + h_3\biggr\}
,\label{equ:ADMM_x2}\\
\mathbf{A}^{(n,m+1)}
:=& \arg\min_{\substack{
            {\mathbf{A}}
        }}
\biggl\{ \frac{\zeta}{2} \Big\|
    \mathbf{A} \cdot \mathbf{\Xi}^{(n,m+1)} - \mathbf{I}
\Big\|^2_F + u_1\biggr\}
,\label{equ:ADMM_A}
\end{align}
where $\mathbf{\Xi}^{(n,m+1)} \triangleq \mathbf{\Xi}\left(\mathbf{x}^{(n,m+1)}_{\text{DD},\text{p,1}},\mathbf{x}^{(n,m+1)}_{\text{DD},\text{p,2}}\right)$ in (\ref{equ:ADMM_A}), and $\mathbf{d}^{(n,m)}$ denotes the ADMM stepsize at the $n^\text{th}$ AO iteration and $m^\text{th}$ ADMM iteration. For the sake of simplicity, $g_1$, $g_2$, and $g_3$ represent the indicator functions associated with constraints (\ref{equ:SP3_C1})--(\ref{equ:SP3_C3}), respectively, with $\mathbf{x}_{\text{DD},\text{p,2}}$ fixed. Similarly, $h_1$, $h_2$, and $h_3$ represent the indicator functions associated with constraints (\ref{equ:SP3_C2})--(\ref{equ:SP3_C3}), respectively, with $\mathbf{x}_{\text{DD},\text{p,1}}$ fixed. The same applies to $u_1$. Note that \eqref{equ:ADMM_x1} remains nonconvex due to a concave part in the objective function and could be solved via SCA.

\section{Numerical Results}
\label{sec:Num_res}
In this section, we verify the DFRC-OTFS performance obtained from the proposed algorithm. In all simulations, the transmit power is set to 30 dBm. The pilot ratio, denoted by $r_\text{pilot}$, represents the proportion of pilot symbols to the total number of transmitted symbols in the DD domain for OTFS, and is defined as $r_\text{pilot} \triangleq \frac{K_p}{K_p+K_c}$. Additionally, $r_\text{GI}$ denotes the proportion of the GI to the total number of grids in the DD domain, and is defined as $r_\text{GI} \triangleq \frac{MN-K_p-K_c}{MN}$. The CP ratio denotes the proportion of the CP length relative to the original OTFS frame, i.e., $r_\text{CP} = \frac{N_\text{CP}}{MN}$.

\begin{figure}[h]   
  \centering
  \subfloat[Spike]{%
    \includegraphics[width=0.3\linewidth]{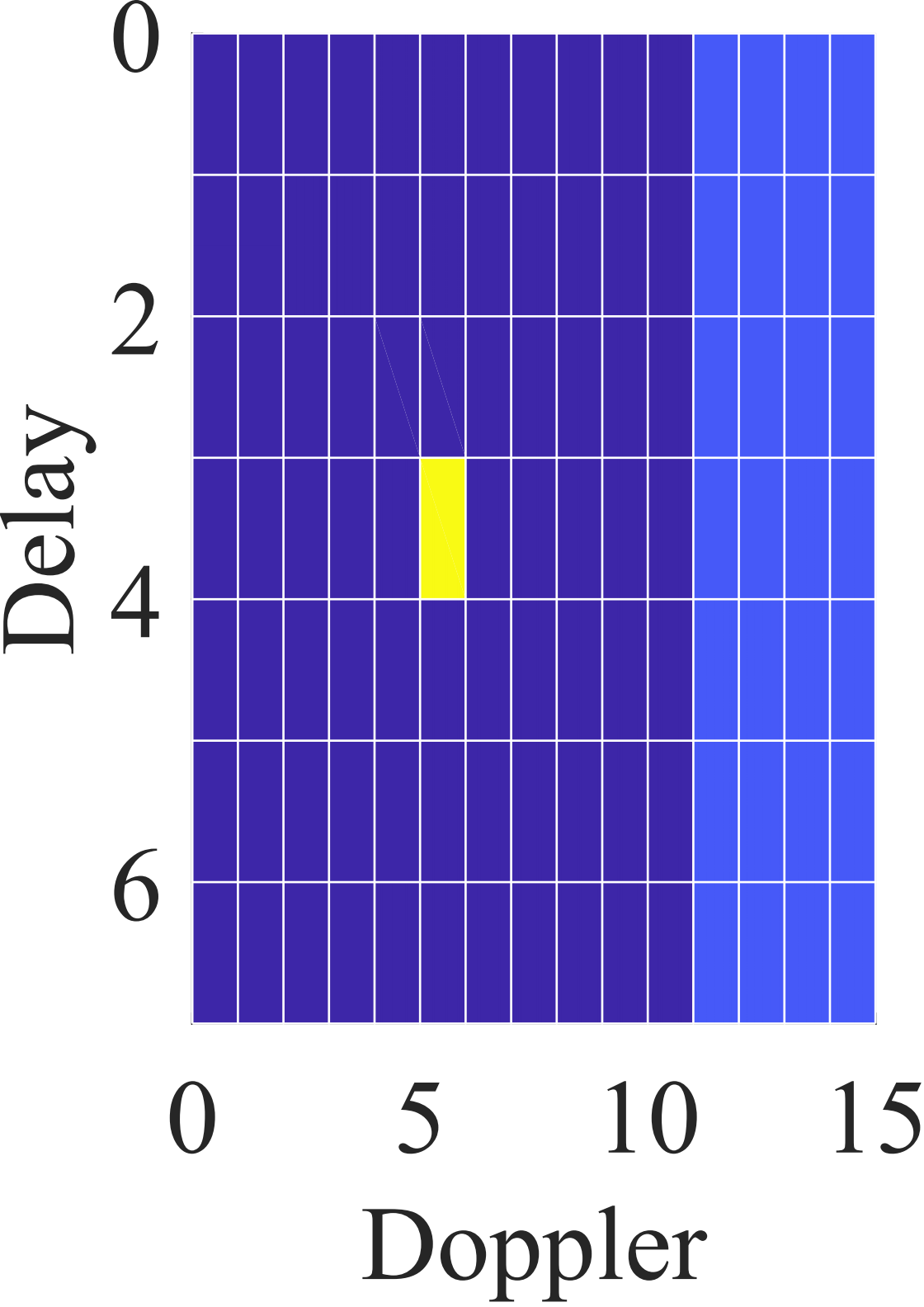}%
    \label{fig:Scheme_spike}%
  }\hfill
  \subfloat[Flat]{%
    \includegraphics[width=0.3\linewidth]{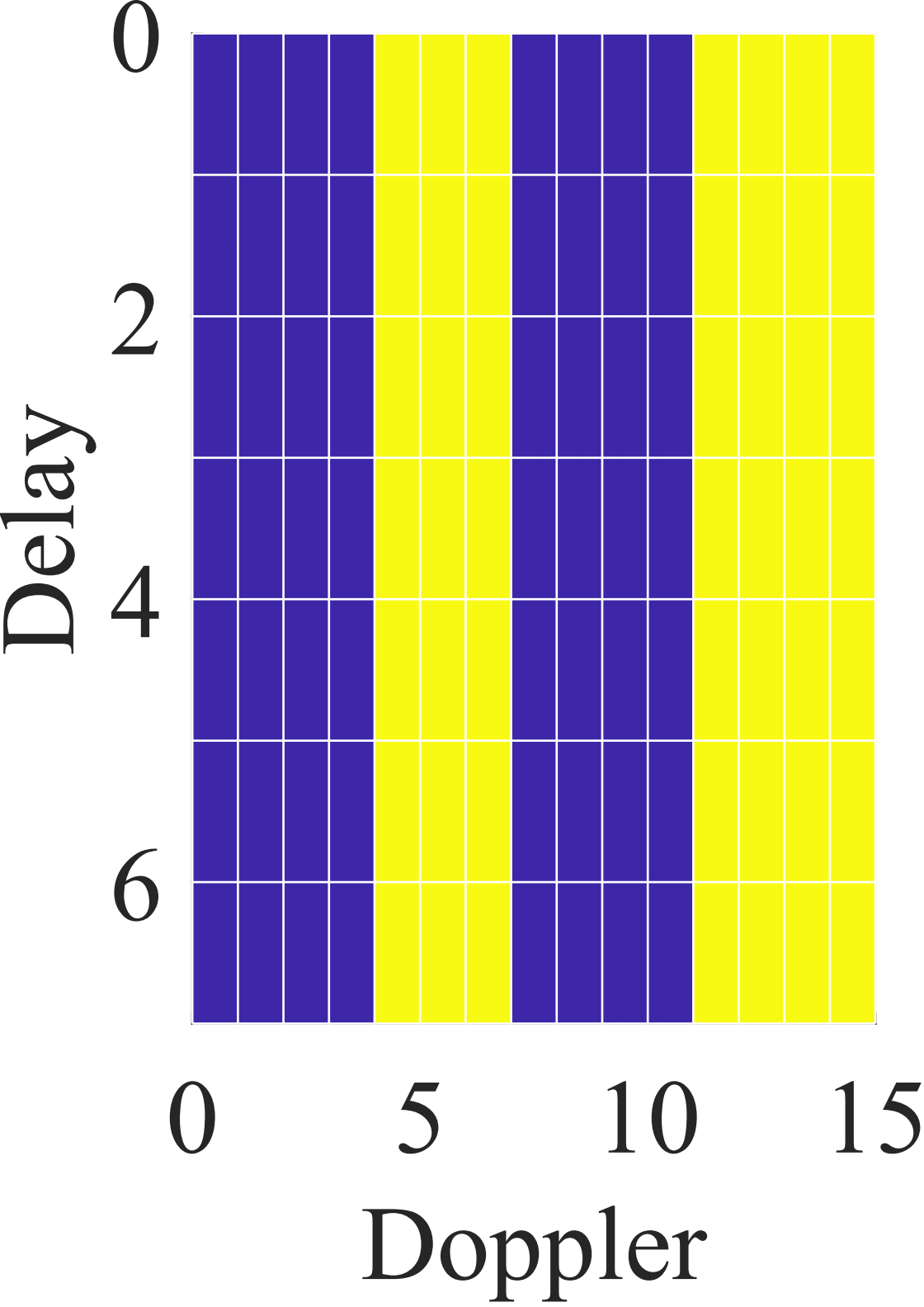}%
    \label{fig:Scheme_flat}%
  }\hfill
  \subfloat[Cluster]{%
    \includegraphics[width=0.3\linewidth]{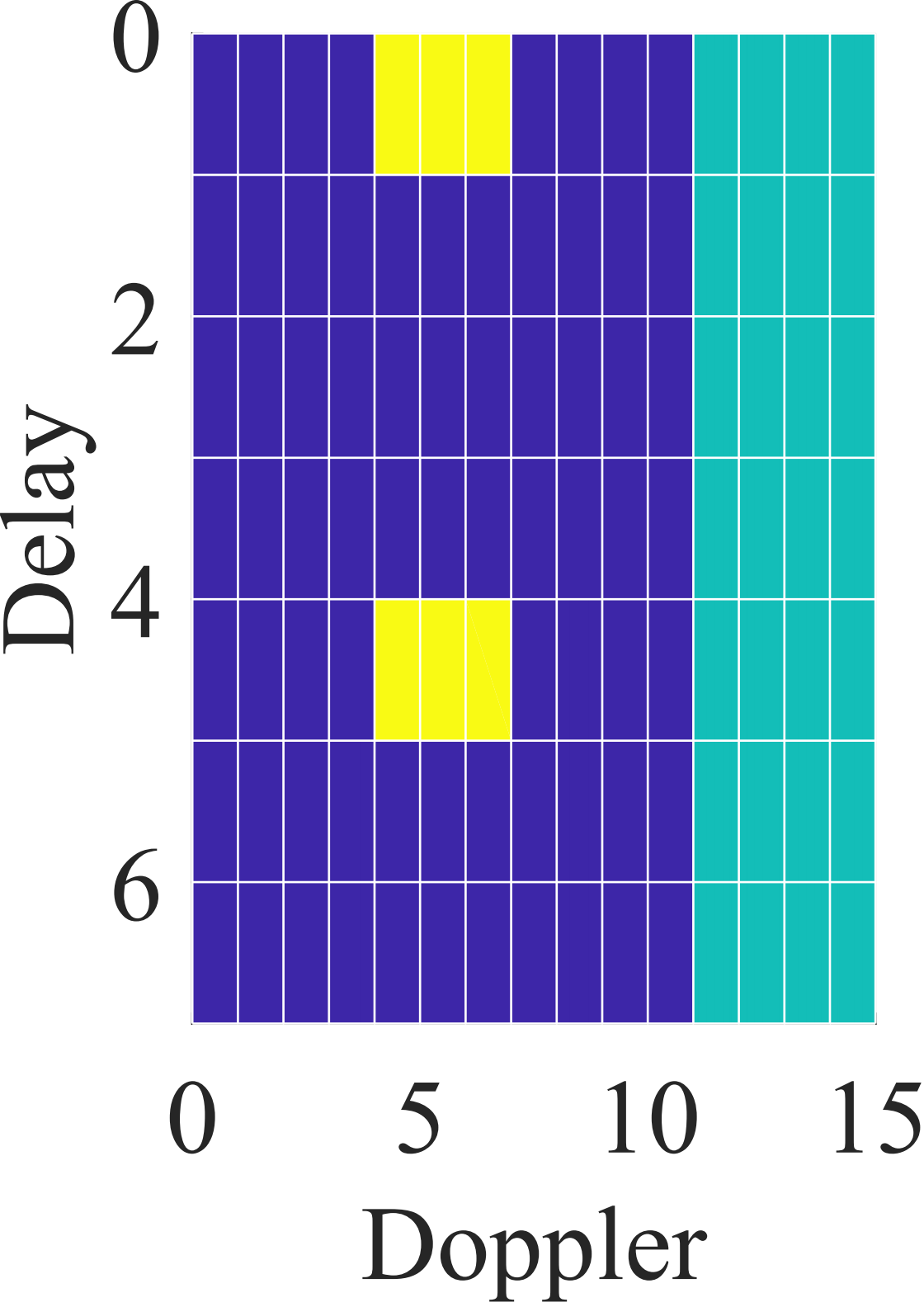}%
    \label{fig:Scheme_cluster}%
  }
  \caption{Spike, flat, and cluster arrangements in the DD domain.}
  \label{fig:Scheme}
\end{figure}

Unless specified, we set the system parameters as follows: the number of subcarriers $M=8$, the number of OFDM slots per OTFS frame $N=16$, the number of data symbols $K_c=40$, the number of pilot symbols $K_p=24$, and the CP ratio $r_{\text{CP}}=0.125$. To mitigate local optima, the optimization has been tested using several starting patterns: spike, flat, and cluster arrangements, as shown in Fig.~\ref{fig:Scheme}. 

\begin{figure}[t]
  \centering
  \includegraphics[width=0.8\linewidth]{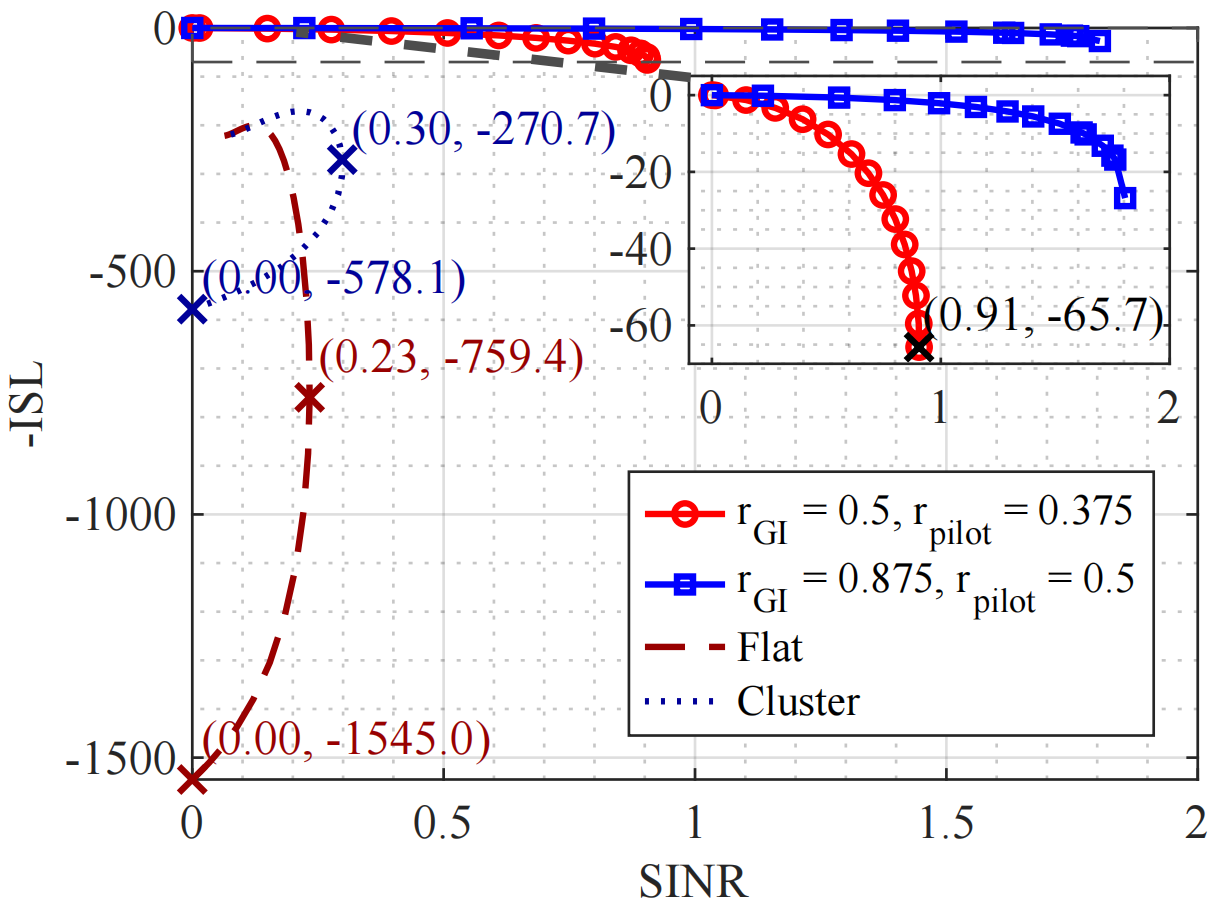}
  \caption{The achievable DFRC performance region (each metric is normalized to its optimal value at $\eta=1$ during optimization, and is displayed on its original scale). Flat and cluster schemes employ $r_\text{GI}=0.5$ and $r_\text{pilot}=0.375$.}
  \label{fig:Opt_Region_c}
\end{figure}
Fig.~\ref{fig:Opt_Region_c} illustrates the sensing-communication performance region for DFRC-OTFS as the DFRC weight \(\eta\) varies from 0 (sensing-centric) to 1 (communication-centric). Fig.~\ref{fig:Opt_Region_c} compares the achievable DFRC performance region of the proposed design with the flat and cluster baselines. The baseline envelopes are traced by sweeping the data-pilot power split from data-only to pilot-only (from top to bottom). Fig.~\ref{fig:Opt_Region_c} demonstrates that the optimized design expands the achievable region relative to both baselines, confirming the effectiveness of the proposed optimization. Specifically, compared with the cluster and flat schemes, the proposed signal achieves ISL suppression gains of 9.44 dB and 13.7 dB (when comparing the worst ISL), respectively, and SINR gains of 4.82 dB and 5.97 dB (when comparing the best SINR), respectively.

\begin{figure}
  \centering
  \subfloat[Zero-Doppler slice
    \label{fig:Zero_Doppler_af}]{
      \includegraphics[width=0.47\linewidth]{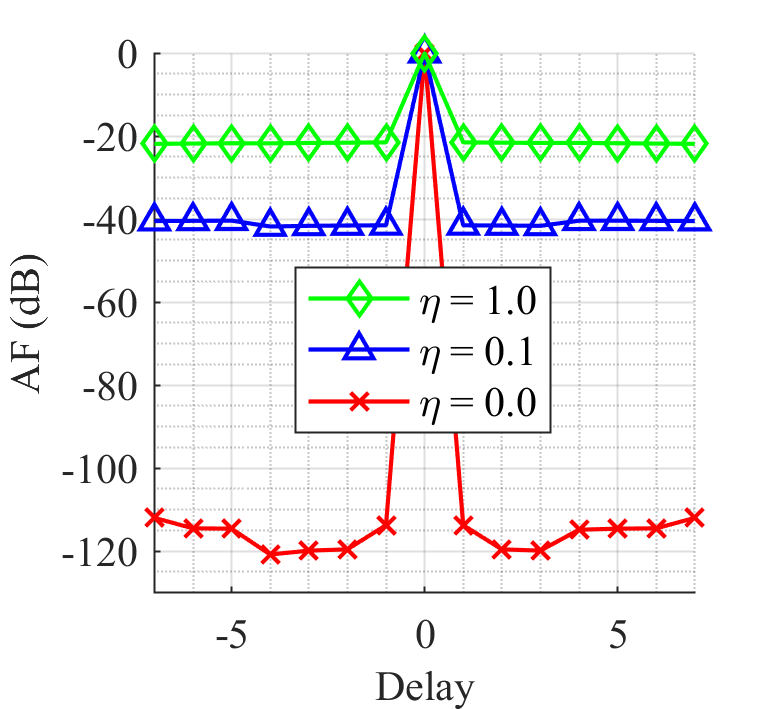}
  }\hfill
  \subfloat[Zero-delay slice
    \label{fig:Zero_Delay_af}]{
      \includegraphics[width=0.47\linewidth]{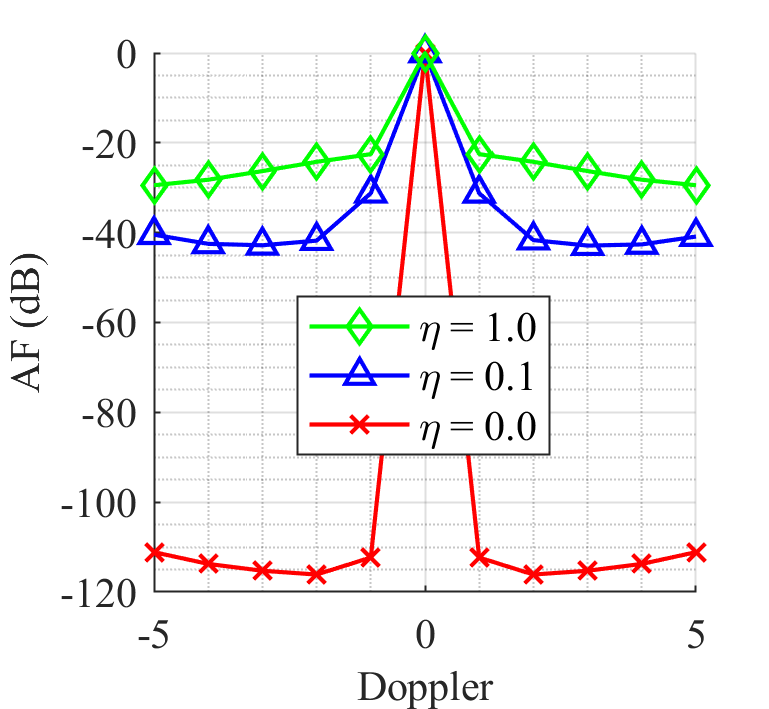}
  }
  \caption{Empirical AFs of optimization results across different ISAC balance parameters to verify our proposed radar metric.}
  \label{fig:Opt_ISL}
\end{figure}
Fig.~\ref{fig:Opt_ISL} shows that the sidelobe level decreases as $\eta$ decreases from 1 (communication-centric) to 0 (sensing-centric), thereby demonstrating the negative effect of random communication symbols on the sensing performance and highlighting the random-deterministic trade-off in DFRC.

\begin{figure}
    \centering
    \includegraphics [width=0.77\linewidth]{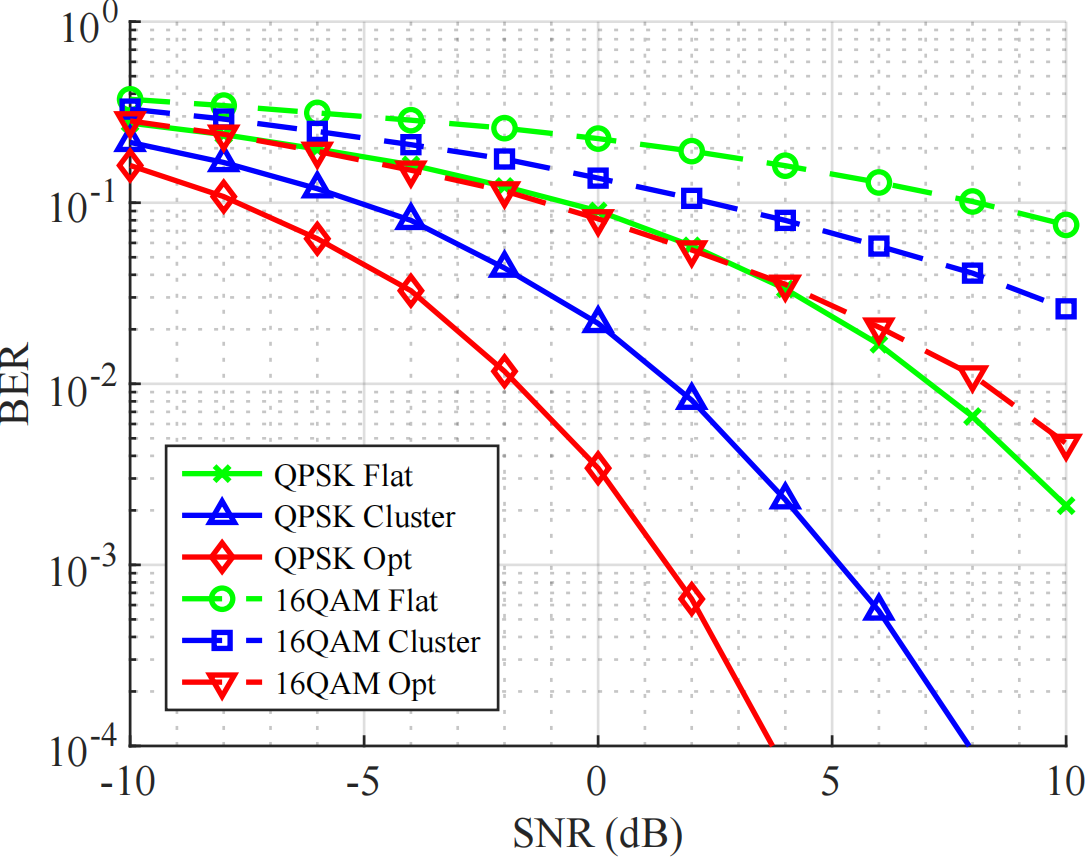}
    \caption{ Monte Carlo results for the optimized, cluster, and flat arrangements with estimated CSI (\(\eta=1\), \(r_{\text{GI}}=0.5\), \(r_{\text{pilot}}=0.375\); \(L=7\), \(Q=3\); \(5\times10^{5}\) trials).}
    \label{fig:BER}
\end{figure}
To further validate the proposed OTFS signal for communication, we simulate the bit error rate (BER) performance versus signal-to-noise ratio (SNR), in comparison with different signal schemes, as shown in Fig.~\ref{fig:BER}. It is evident that the optimized signal achieves the lowest BER, followed by the cluster arrangement and the flat arrangement, for both QPSK and 16-QAM modulation schemes. The BER simulation in Fig.~\ref{fig:BER} also demonstrates the effectiveness of our derived capacity lower bound as the communication metric.

\section{Conclusion}
\label{sec:conclu}
This paper studied DFRC signal design for OTFS-based ISAC systems. We first presented an OTFS system model and derived two metrics: (i) a tractable channel capacity lower bound that accounted for channel estimation error, whose dominant term (SINR) served as the communication metric; and (ii) an ISL metric for sensing that jointly captured deterministic pilots and random data. The joint design was formulated and solved via an AO framework combining SCA and ADMM. Simulations showed that the proposed method significantly expanded the DFRC region for OTFS, compared with conventional schemes. The proposed metrics were validated using ISL and BER results.

\bibliographystyle{IEEEtran}
\bibliography{IEEEabrv,mybib}

\end{document}